\documentstyle[12pt,aasms4,psfig]{article}

\def\kms{\ifmmode \hbox{km~s}^{-1}\else km~s$^{-1}$\fi}
\def\etal {{\it et al.}}
\slugcomment{revised AJ version: \quad {Aug 12, 2004} }

\begin{document} 

\hyphenation {non-re-la-ti-vi-stic}
\def\iacen{\'{\i}}

\title{CO in the Bipolar-Radio-Continuum Galaxy NGC 3367}

\author{J. A. Garc\'{\i}a-Barreto\altaffilmark{1}, N.Z. Scoville\altaffilmark{2}, 
J.Koda\altaffilmark{2}, \& K. Sheth\altaffilmark{2} }
\altaffiltext{1}{Instituto de Astronom\'{\i}a,
Universidad Nacional Aut\'onoma de M\'exico, Apdo Postal 70-264, M\'exico D.F.
04510, M\'exico; tony@astroscu.unam.mx}
\altaffiltext{2}{Caltech, Radio Astronomy Millimeter Group, 1200 E. California Blvd,
MS 105-24, Pasadena, California 91125, USA}

\begin{abstract}
\noindent 

	CO(1-0) emission has been imaged at 2\arcsec ~resolution in the central 10 kpc of the barred spiral 
	galaxy NGC 3367. This galaxy has bipolar synchrotron lobes out to a 
	radii 
	of 6 kpc, straddling the compact nucleus. The peak molecular emission is 
in a source of radius 2\arcsec (425 pc) centered on the galaxy nucleus. 
The molecular gas mass is $\sim3\times10^8 M_{\odot}$ in this 
peak and $\sim5.9\times10^8 M_{\odot}$ within a radius of 
$4.5$\arcsec (950 pc). The very large gas masses in the central source 
imply extinctions sufficiently high to 
completley obscure optical emission lines (e.g. BLR) 
associated 
with the nuclear radio source. 
The observed Balmer lines probably originate in the narrow line region 
few hundred pc from the 
nucleus. The CO emission in the central region is elongated  
NE - SW, very similar to the position angle of the large-scale synchrotron lobes. 
This elongation is likely due to the non-axisymmetric gravitational 
potential of the stellar bar. 
We infer that the NE radio continuum lobe is on the far side of the 
galaxy and the SW lobe is on the near side. The 
central mass of molecular gas is sufficient mass to power the AGN accretion 
luminosity for over 10$^8$ yrs at 3 M$_{\odot}$ yr$^{-1}$. 
\keywords{galaxies: structure --- galaxies: spiral --- galaxies: individual (NGC 3367) 
--- ISM: molecules}
\end{abstract}

\section{INTRODUCTION}

	Multiwavelength imaging and spectroscopy are vital to understanding 
	the 
	feeding and observed characteristics of active galactic nuclei (AGN). 
	Observations of the molecular gas using the CO (J=1-0) emission line 
	are especially important since this gas may be the fuel for the AGN.
	The CO (J=1-0) emission 
peaks toward the centers of most spiral galaxies (\cite{sak99b}) and the 
associated interstellar dust and its extinction often determine the 
emergent optical-UV 
	spectrum of the AGN.  In the optical, 
galactic nuclei are classified according to the width of Balmer lines, emission of high ionization atoms 
and the ratios of permitted to forbidden emission lines; in the radio 
continuum, the activity can manifest as central triple sources  
(\cite{col96,ho01}, \cite{wil87}, \cite{cra92}) or circumnuclear radio continuum 
structures (e.g. NGC 1326 \cite{gar91a}, NGC 4314 \cite{gar91b}).

	NGC 3367 (at 43.6 Mpc) is a barred spiral (SBc) galaxy showing a 
	remarkable triple 
radio continuum synchrotron structure: a central unresolved (d$\leq 65$ pc) source and two larger lobes straddling 
the nucleus at a radius of about 6 kpc projected on the sky (\cite{gar98,gar02}). It 
also shows an unresolved H$\alpha$+N[II] central source (diameter $\leq 
5$\arcsec or $\leq$ 1 kpc). It has a 
modest far infrared luminosity L$_{FIR}\sim 2\times10^{10} L_{\odot}$ 
(\cite{soi89}) and is a source of X-ray 
emission with  L$_X\sim 9\times10^{40}$ ergs s$^{-1}$ (\cite{fab92}). 
Optically, it is classified as a weak Liner or HII 
nucleus (\cite{ver86,ho95,ver97,ho97}). The triple synchrotron source suggests that there must be an AGN and the 
interstellar medium (ISM) may exhibit the effects of this energy release 
from the  central AGN. The orientation of the 
central synchrotron source and the two lobes at PA$\sim35^{\circ}$ (see Fig. 11 this paper or Fig. 3 
\cite{gar98}) indicate that the 
radio jet outflow is not strictly parallel to the rotation 
axis of the galaxy disk (PA$\sim140^{\circ}$ for the kinematical minor axis \cite{gar01}). 

The HI gas mass for NGC3367 is $\sim 7\times10^9 M_{\odot}$ (\cite{huc89}). 
No previous observations of the molecular gas have been reported; 
however, from the CO-FIR correlation (\cite{you86}) one  
expects a total molecular gas of $\sim1.6\times 10^9 M_{\odot}$. H$\alpha$ images show emission 
from the disk of the galaxy, indicating several 
regions of massive OB star formation in the disk (\cite{gar98,gar01}). 
CO interferometric observations of spiral galaxies have indicated 
higher concentrations of molecular 
gas in the centers of barred spirals as compared to normal spiral galaxies, supporting the idea of 
bar-driven gas transport (\cite{sak99a,sak99b,sak00,kod02,she02,she04}). 

	In this paper we report CO(1-0) interferometric mapping of NGC 3367. 
	The images reveal a central molecular 
gas concentration with radius 4.5\arcsec (or 0.9 kpc) and mass $\sim5.9\times 10^8 
M_{\odot}$. This gas concentration is unresolved at our highest CO angular 
resolution ($2.6 \times 1.9$\arcsec or 550 pc $\times$ 400 pc), suggesting that H${\alpha}$, CO and 
plasma (non-thermal and thermal) are co-extensive in the central 450 pc. 
The CO is elongated 
in a SW - NE direction parallel to the radio continuum emission. The 
morphology of the synchrotron emission suggests a connection between the central source and 
the large scale radio lobes; however, the CO gas is very likely in the disk of the galaxy
while the radio lobes are likely to be above and below the galaxy disk. 
In $\S$2 we describe the observations; results are presented in $\S$3; 
analysis and discussion are presented in $\S$4; finally conclusions are presented in $\S$5.

\section{OBSERVATIONS}

	The central disk of NGC 3367 was observed using the Owens Valley Millimeter Array 
(\cite{sco93}) in the CO (1-0) transition in 4 antennae 
configurations (C, L, E, H) from 2003 September to 2003 December. The array consists of six 10.4 m radio 
telescopes and the baselines ranged from 20 to 242 m. The synthesized 
beams (FWHM) were 
$4''.2\times2''.7$ (natural weighting) and $2''.6\times1''.9$ (uniform weighting). Spectral 
resolution was provided by a digital correlator configured to 128$\times$ 2MHz 
(5.2 \kms ~resolution). The nearby quasar 1058+015 (2.54 Jy at 115 
GHz) was used to calibrate the phase and gain variations. Twenty channel maps, each of 15.6 km s$^{-1}$ width, 
were cleaned and then combined for the moment maps. Our 1 $\sigma$ rms detection was 0.15 Jy/\kms (uniform 
weighting) and 0.2 Jy/\kms (natural weighting).
  
\section{RESULTS}

	Figure 1 shows the interferometric spectrum integrated over the 
	central 
	17.5\arcsec ($\sim1.8$ kpc) from the natural weighting data (fwhm 4$''.2 \times 2''.7$). The CO (1-0) channel maps in 
	15.9 \kms ~bins is
	shown in 
Figure 2 superimposed on an I-band image (centered at 8040~\AA). CO emission is detected most strongly
in the 
central region of the galaxy -- a compact source with elongated to the eastern and western sides (perhaps 
along the  dust lanes in the stellar bar); weaker CO emission is also 
seen associated with the spiral arm on the east side of the stellar bar and from the western end of the stellar bar. 
At the last position there is no radio continuum synchrotron emission (see Fig. 6 
of \cite{gar98}) but there is a weak H$\alpha$+N[II] emission feature(see Fig. 5 of \cite{gar98}). 

\subsection{The Nuclear Region}

Figures 3 and 4 shows the total CO line flux 
integrated over 300 \kms ~,superposed on the H${\alpha}$+[NII] (continuum 
subtracted) and $I$ band images (\cite{gar96,gar96b}). 
The strongest CO emission (65 $\sigma$) arises at the compact 
nucleus and from a circumnuclear structure 
(see below). The peak intensity of the central emission, in the natural weighting spectrum, is about 250 mJy and the velocity 
width is 80 \kms (FWHM). The  
azimuthally averaged radial distribution of the CO indicates a radius 
(out to our detecion threshold) of 
$\sim$7\arcsec (1.5 kpc). A similar, unresolved CO structure was 
detected in the center of NGC 3368 (SAB(rs)ab, \cite{sak99a}).  The deconvolved 
size of the central source extent is 525 pc $\times$ 300 pc. 

\subsection{CO and Radio Synchrotron Emission}

	Figure 5 shows a zoomed image of the central CO concentration from the high resolution image. 
The {\it unresolved}, peak emission is at $\alpha(J2000)=10^h46^m34^s.948, 
\; \delta(J2000)=+13^{\circ}45'02.7$\arcsec. 
This position agrees closely with the peak of the synchrotron radio continuum emission 
($\alpha(J2000)=10^h46^m34^s.956, 
\;\delta(J2000)=+13^{\circ}45'02.94$\arcsec [\cite{gar02}]). A secondary 
peak occurs to the SW at $\alpha=34^s.66; \; 
\delta=45'00.90$\arcsec. The 20-cm radio continuum from the central region 
has an extension towards 
PA$\sim225^{\circ}$ connecting to the southwest lobe (see Figs. 11 and 12).
This is very similar to the PA of the secondary CO peak; however no radio 
synchrotron emission was actually detected at the secondary CO peak 
and it is likely that the radio jet is above the galactic disk at 
this radius.

\subsection{CO in the Bar and Disk}

Extended CO emission 
is detected in the central region of NGC 3367 at PA NE - SW, 
along the center of 
the stellar bar. Weak extended emission is observed in the CO velocity maps from the west side of disk (blue 
shifted at 3006 \kms to 3054 \kms ~,see Fig. 2). Those are the expected locations of the leading edge dust lanes from a bar rotating in the 
counter-clockwise direction (assuming trailing spiral arms). 

The north-south structure on the western side of the 
galaxy is detected at 5$\sigma$ (Figs. 3 and 
	4);  all of the other extra-nuclear emission features are less than 5 $\sigma$ 
	(1 $\sigma=200$ mJy \kms). Careful examination of the optical $I$ band 
photometry and the CO contours (Figs. 3 and 4) indicates that only the 
weak, unresolved source in the western side of the disk 
could be associated with the stellar bar; the north-south CO structure 
on the west side is most 
likely not related to the stellar bar. The N-S CO structure is 
probably associated with an H$\alpha$ emission region (Fig. 3).

\subsection{Gas Kinematics and Dynamical Mass}

	Figure 6 shows the CO centroid velocities in contours superposed on 
	the integrated CO image in grayscale. The systemic velocity of 
the galaxy is $\sim3032$ km s$^{-1}$ from H$\alpha$ Fabry-Perot (\cite{gar01}). 
The red shifted gas lies to the NE and the blue 
shifted gas to the SW. This polarity is the same as that seen in H$\alpha$ Fabry Perot observations at larger distances from the nucleus 
 --  the NE side receeding; the SW approaching. The PA of 
the receding semi major axis (from H$\alpha$ Fabry Perot) 
is 51$^{\circ}$ (\cite{gar01}), indicating that the northwest is the near 
side of the galaxy (assuming 
trailing spiral arms). For circular rotation, the contour at the 
systemic velocity should be a straight line at 
PA=$140^{\circ}$; instead, the contour at 3038 \kms (Fig. 6) is 
clearly curved. This twisting of the constant velocity contours is 
commonly seen in barred spiral galaxies, due to the non-circular 
motions associated with highly eccentric orbits.

The velocity dispersion (2nd moment) map is shown in Figure 7, derived 
from the naturally weighted map. Average velocity dispersions are $\sigma_v\sim15$ km s$^{-1}$; 
however, two regions show extraordinarily large dispersions : at PA$\sim-40^{\circ}$ (corresponding 
to the kinematic minor axis) where $\sigma_v\sim40$ km 
s$^{-1}$ and southwest of the nucleus where the dispersion reaches 75 km 
s$^{-1}$ ($\alpha=34^s.66;\delta=45'00''.90$). The latter region is 
particularly bright in the CO integrated intensity map (Fig. 5) and 
corresponds to a location where there is very weak synchrotron emission (see Fig. 2 in \cite{gar02}). 

Figure 8 shows a position-velocity (PV) diagram from the central region 
along the major axis (PA = 51$^{\circ}$) of the galaxy. The total velocity dispersion at the center is nearly 130 km s$^{-1}$. 
The blue peak in Figure 8 is at 3022 km s$^{-1}$ and the red shifted peak is at 3068 km s$^{-1}$. The peaks 
arise from positions about 50 to 60 pc from the central maximum of the CO integrated emission. If the gas 
were rotating on the plane of the disk of the galaxy (at r$\sim175$ pc or $0''.8$ from the center) at about 65 
km s$^{-1}$ / [sin$(i)$]=130 km s$^{-1}$, the implied dynamical mass would be M$_{dyn}\sim RV^2/G =7\times10^8$ 
M$_{\odot}$ (assuming $i\sim30^{\circ}$, \cite{gar01}). This dynamical 
mass is low compared to similar estimates at R $\leq$ 500 pc for other barred galaxies 
(For type SAB's , \cite{sak99b} find only 1 out of 17 having 
equivalently low dynamical mass.)

\section{DISCUSSION}

\subsection{Molecular Gas Mass}  

From the integrated CO line fluxes we have estimated the molecular 
gas mass in central regions of NGC 3367  using the  a standard Galactic 
conversion factor ($\alpha_{Gal}=4 M_{\odot}$ (K \kms pc$^2$)$^{-1}$ 
[\cite{you82}]). The molecular gas mass M(H$_2$), in units of solar masses, 
is then obtained using M(H$_2)=1.5 \times 10^4 (D_{Mpc})^2 
S_{CO} (\alpha/(\alpha_{Gal})$ 
(Jy \kms), including a factor of 1.36 to account for He within the 
molecular gas (\cite{sak99a}). 

It is of course possible that the standard Galactic CO-to-H$_2$ conversion factor is not appropiate to the 
central region of NGC 3367 where there could be a starburst and the 
physical conditions would be different from those in a  
Galactic GMC. CO is 
optically thick and the conversion factor is roughly proportional to $\langle n_{H_2} \rangle ^{1/2} / T_{ex}$ 
(\cite{dow93,sco97}). 
To some extent, the density and temperature dependences are likely to 
offset each other since where the gas density is high, there will be 
elevated star formation rates and higher temperatures.
Our observations only involve one transition;  we therefore have no constraints 
on the CO excitation, and thus adopt the Galactic $\alpha$. 

	From the high resolution map (Fig. 5), the peak has a deconvolved size of 2$''.5 \times 1''.45$ 
(525 pc $\times$ 300 pc), an intensity of 11 Jy \kms ,  implying a molecular 
gas mass $3\times10^8 M_{\odot}$. 
The central CO emission, from the low resolution map (Figs. 3 and 4), is slightly resolved and elongated. The molecular 
mass estimate within a radius of $4''.5$ (950 pc) is 
$5.9\times10^8 M_{\odot}$. Lastly, using the total flux in the 
naturally weighted map out to a radius of 27$''$ (5.7 kpc), we 
obtain $2.6\times10^9 M_{\odot}$. This last estimate  is a lower limit 
to the total gas mass in the galaxy since the interferometer will resolve 
out extended spatially components.  

\cite{sak99b} observed the central regions of a sample of 20 spiral 
galaxies and found a range for M(H$_2$) of $3.3\times10^7$ M$_{\odot}$ to 
$8.1\times10^8$ M$_{\odot}$ inside 500 pc radius.
Average molecular masses inside a radius of 500 pc as a 
function of galaxy  type are: 
$4.9\times10^8$ M$_{\odot}$ for SB+SAB galaxies; $2.1\times10^8$ M$_{\odot}$ for SA; 
$3\times10^8$ M$_{\odot}$ for HII galaxies; $3.3\times10^8$ M$_{\odot}$ for transition galaxies and 
$3.4\times10^8$ M$_{\odot}$ for Seyfert galaxies (\cite{sak99b}). Thus 
the molecular gas mass for the 
central region of NGC 3367, $5.9\times10^8$ M$_{\odot}$, is very similar 
to that of HII and 
transition spiral galaxies but it is high for a Hubble type SBc galaxy 
(\cite{sak99b}).

\subsection{Obscured Optical Emission from the Compact Nucleus}

The molecular gas surface density is $\Sigma(H_2)=2.5 \times 10^3$ 
M$_{\odot}$ pc$^{-2}$ at the peak (within 
525 pc $\times$ 300 pc) 
and 210 M$_{\odot}$ pc$^{-2}$ for the central 1.9 kpc region. For comparison, 
the surface density is 1500 M$_{\odot}$ 
pc$^{-2}$ inside 715  $\times$ 1600 pc in NGC 1530 
(\cite{dow96}). And in  two early type SBa galaxies with 
circumnuclear rings (NGC 1326 and NGC 
4314), $\Sigma(H_2)=580 M_{\odot}$ pc$^{-2}$ and 
$500 M_{\odot}$ pc$^{-2}$ respectively (\cite{gar91a,gar91b}).  

	The high value of $\Sigma(H_2) (2.5 \times 10^3$ M$_{\odot}$ pc$^{-2}$ within 262 
$\times 150$ pc  in 
NGC 3367) is equivalent to an average column density of $\sim1.5\times10^{23}$ cm$^{-2}$. 
This gas column translates into a visual 
extinction towards the compact nucleus of  $\sim$150 mag (using 
N(H$_2$)/A$_v=0.94\times10^{21}$ cm$^{-2}$ mag$^{-1}$, \cite{you82}). 
On the other hand, optical spectra of NGC 3367 show 
moderately broad H$\alpha$+[NII] line emission with FWHM velocities  
up to 600 km s$^{-1}$. The H${\beta}$ line intensity 
is about twice that of the [O III] $\lambda 5007$ line and there are possible Helium lines ($\lambda 
4686$) indicative  of W-R stars (\cite{ver86,ho95}). The high extinction 
associated with the molecular gas would heavily obscure any H$\alpha$ 
emission from the compact nucleus, suggesting that the observed H$\alpha$ emission must originate from the ionized gas in the narrow line region
outside the dense molecular gas. The large dust opacity would explain 
the lack of wide Balmer lines. The 
observed optical spectrum from NGC 3367 
(\cite{ver86,ho95,ver96}) is consistent with the ionized gas emission lines 
arising from the surfaces of the 
molecular clouds. 

\subsection{The Nuclear Gas Distribution}

	The high resolution CO map of the central region (Figs. 3, 4, 5) indicates a very striking distribution 
of the molecular gas as compared to the radio continuum distribution. The 
CO maps show extended emission to the west and east near extensions in the radio continuum 
synchrotron  (see Figs. 10 and 11). The PV diagram (Fig. 8) shows two peaks 
along the major axis at different velocities. There are several 
possible explanations for these two peaks in the CO, 
straddling the nucleus : (a) a central hole in the gas, (b) an 
increase in the line broadening on the nucleus or (c) a decrease in the  
CO emissivity due to extreme physical conditions there. 

In the case of 
NGC 3367 there is additionally a plasma outflow from the AGN which 
might clear gas from the vicinity of the AGN. The orientation of the synchrotron triple sources 
and the plasma outflow is likely to be perpendicular to the accretion 
disk -- and both are apparently inclined 
with respect to the galaxy disk (\cite{pri99,kin00}). It is unlikely 
that the outflowing plasma is directed by interaction with the dense molecular gas (\cite{hen81,bri84,bla90}). 
Molecular gas to the NE 
($\alpha=35^m.15; \; \delta=45'03''$) is in a region of no detectable radio emission (in fact the 
radio continuum synchrotron emission shows a sharp cutoff [see Figs. 
10 and 11]).  
The kinetic energy required to significantly displace the 
molecular gas in the west source is $\sim 1/2 mv^2 \simeq 2\times10^{52}$ 
ergs, assuming a velocity of about 15 \kms and a mass of about 10$^7 
M_{\odot}$. 

Based on the available observations (optical red continuum, $H\alpha$, radio continuum synchrotron, 
CO(1-0)), we favor the interpretation that the central density of molecular 
gas is reduced compared to that at $\sim 100$ pc. NGC 1068 is  another spiral galaxy with bipolar synchrotron 
lobes (although smaller in extent, only $\sim450$ pc). High angular resolution 
CO(1-0) and CO(2-1) interferometry of NGC 1068 reveals two 
concentrations molecular gas straddling the compact nucleus at a PA perpendicular 
to that of the plasma 
outflow within 100 pc (\cite{sch00}). In the case of NGC 1068, 
Schinnerer \etal  ~suggest a 
model with the molecular gas in a warped disk of thickness 10 pc and a radius of about 150 pc. 
The BLR in NGC 1068 would be obscured to our line-of-sight; however,
optical polarization studies do indicate its existence in broad 
polarized line wings on the H$\alpha$ 
emission (\cite{ant85}).

\subsection{Gas Flow to the Nucleus}

In barred spiral galaxies,  exchange of angular momentum of bar and gas (molecular clouds) cause gas flow inwards 
from the corotation radius (CR) to the inner Lindblad resonance, forming a circumnuclear structure or ring (\cite{but86,but96}).
Gas should also be driven outwards from CR to the outer Lindblad 
resonce (OLR).  Inside the CR, the gas moves on $x_1$ orbits elongated 
parallel to the bar and on $x_2$ orbits perpendicular to the bar 
(\cite{con89}). Near the ILR the orbits change from $x_1$ to $x_2$; viscous 
dissipation forces a continuous change 
between the two families of orbits and shocks can result. Barred galaxies like NGC 1530 (SBb) shows CO gas 
from the central region [M(H$_2)=6 \times 10^9$ M$_{\odot}$] as well as from along the dust lanes along the bar 
(\cite{rey97,rey98}). On the other hand, barred galaxies like NGC 4314 (\cite{gar91b,reg99}) or NGC 5383, show no 
molecular gas in the dust lanes except where the dust lanes join the nuclear circumnuclear structure (\cite{reg99}). 

 As stated earlier, the 
amount of total mass (M(HI) + M(H$_2$)) is about 9$\times10^{9} M_{\odot}$ in 
NGC 3367, which is similar to the amount of 
mass found in other spiral galaxies (\cite{sak99a,sak99b,sak00,she00,kod02}) 
yet NGC 3367 shows two large synchrotron lobes at either side of the 
center (at radius 6 kpc from the nucleus); a very similar outflow is observed in NGC 3079 [\cite{kod02}]). 
No broad line region is observed in NGC 3367, very likley the result 
of high extinctions in the molecular gas. The optical disk of 
NGC 3367 is peculiar (\cite{gar96}): (a) there is a bright optical structure 
forming a semi-circle at a 
radius of approximately 10 -- 11 kpc SW of the center; this structure is bright in optical 
red continuum as well as in narrow line H$\alpha$+[NII] images; (b) it 
has spiral arms apparently joining this semi-circle structure 
(to the NW) at an angle larger than 45$^{\circ}$; (c) the NW spiral arm 
seems to consist of  
two parallel structures (both in red images and in H$\alpha$+[NII] images); (d) 
the north-east (NE) end of the stellar bar exhibits very large velocity 
dispersions in H$\alpha$ \cite{gar01}. Our previous 
analysis of the optical morphology suggested a ringlike density wave 
propagating outwards possibly the result of a collision of NGC 3367 with 
a smaller galaxy (\cite{gar96}). The perturber might  
have traveled approximately perpendicular to the disk of NGC 3367 in an 
off-center collision. 
Simulations of such  galaxy collisions indicate an arclike density wave which propagates outwards from the point 
of impact (\cite{lyn76,too78,mih94,ger96}). In contrast to central 
collisions, the off-center collisions create 
elongated (non-circular) structures (\cite{too78,ger94}).  Another possible scenario for 
NGC 3367 might be a merger with a dwarf galaxy (\cite{her89}) in 
which the dwarf is gradually tidally disrupted and it's remains are distributed 
asymmetrically in the disk by differential rotation. 

The {\bf total} star formation rate can be crudely estimated from 
the bolometric luminosity using : SFR$\sim 8\times10^{-11} L_{TOT}/L_{\odot} 
M_{\odot} yr^{-1}$ (\cite{sco83}). For the observed luminosity of 
L$_{TOT}=L_B+L_{FIR}\sim6.8\times10^{10} L_{\odot}$, we obtain SFR$\sim 5 M_{\odot} yr^{-1}$
which is similar to that of the Galaxy.

\subsection{Other Galaxies with Similar Nuclear Characteristics}

Other barred galaxies show bipolar synchrotron emission from their nucleus. Most notably, NGC 1068, 
(SBb, classified as Sy 2, and hidden Sy 1) has a synchrotron bipolar outflow from the nucleus  
$\sim950$ pc in diameter (\cite{van82,ulv87}). NGC 3079 (SBc) has a synchrotron 
bipolar outflow out to a 
radius of $\sim1.2$ kpc perpendicular to the plane of the disk (\cite{dur88,kod02}). NGC 5548 
(S0a, classified as Sy 1) exhibits strong and variable X-ray emission and bipolar synchrotron 
emission; the bipolar emission diameter is only 4 kpc (\cite{ho01}). Deep optical images of 
NGC 5548 show tidal tails suggesting a past merger (\cite{tys98}). NGC 5643, SABc (Sy 2), has a radio 
continuum and H$\alpha$ morphology  elongated at the PA of the stellar bar. The radio 
continuum extension in NGC 5643 is about 1.7 kpc in diameter and the 
galaxy has no companions within 10 
D$_{25}$. 

\section {CONCLUSIONS}

	We have carried out interferometric (mm OVRO) observations of the 
CO (1-0) emission from the radio-bipolar lobe galaxy NGC 3367. The observations indicate that the molecular gas is concentrated in the 
central (9$''$) region with some weak emission from regions just beyond the ends of the (optical) stellar 
bar to the W and E sides. 

The molecular gas mass is 3$\times10^8 M_{\odot}$ in the central peak of 
the emission ($2.5\times 1.45$\arcsec or 525 $\times$ 300 pc)
and $5.9\times10^8 M_{\odot}$ 
within a radius of $4''.5$ (950 pc). The high molecular gas mass in the center of the galaxy 
translates in a extremely large optical depth which is likely to block 
optical broad line emission from the AGN. 

	Our high angular resolution ($2.6\times1.9$\arcsec ~,FWHM) map of the central emission indicates a unresolved 
source at $\alpha=10^h46^m34^s.948, \delta=+13^{\circ}45'02''.7$ which 
coincides with the nuclear radio continuum 
peak ($\alpha=10^h46^m34^s.956, 
\delta=+13^{\circ}45'02.94$\arcsec). In addition, we have detected 
emission extended from the central peak to the NE and SW. 
This is also the direction of 
the arc second extended synchrotron emission leading to the 12 kpc diameter lobes.

	The velocity field of the central region indicates two peaks which 
	we believe reflect the existenece of a density decrease with the 
	central 100 pc around the AGN. This clearing might result from 
	either the AGN winds or more likely the gas dynamics at the inner 
	Lindblad resonance.

\section*{Acknowledgements}

	 We thank the staff of the Owens Valley Radio Observatory. The Owens Valley Millimeter Array is supported by 
the National Science Foundation (NSF) grant AST 99-81546. JAG-B would like to thank Nick Scoville and the Radio 
Astronomy Millimeter Group at Caltech for their hospitality during a six month visit where part of this paper was 
written and to DGAPA-UNAM (Mexico) for its partial financial support during his stay at Caltech. JAG-B would also 
like to thank the time allocating committee of the OVRO millimeter array for the allocated observing time for this 
project.

\clearpage
\newpage
\begin{table}
\small
\caption[ ]{Barred Spiral Galaxy NGC 3367}
\begin{flushleft}
\begin{tabular}{lcr}
\hline
Property  &     & Reference \cr
\hline
Hubble Type & SBc(s) & 1 \cr
Right Ascension (J2000) & 10$^h$ 46$^m$ 34$^s$.956 & 2 \cr
Declination (J2000)     & +13$^{\circ}$ 45$'$ 2$''$.94 & 2 \cr
Distance    & 43.6 Mpc & 3 \cr
Optical Diameter & 2'.5 & 3 \cr
Log L$_B$        & 10.68 & 3 \cr
Optical Classification & HII, Sy 2-like & 4 \cr
M(HI)              & 7$\times10^9 M_{\odot}$ & 5 \cr
V$_{Systemic}$(HI) & 3030 \kms  & 5 \cr
Far Infrared Luminosity (IRAS) & 2.1$\times10^{10} L_{\odot}$ & 6 \cr
Star Formation Rate & 5 & 7 \cr
Position Angle Disk & 109$^{\circ}$  & 8 \cr
Dust Temperature (IRAS) & 34$^{\circ}$ K & 9 \cr
Stellar Bar diameter & 32$''$ (6.7 kpc) & 9 \cr
PA Bar & 65$^{\circ}$  & 9 \cr
Inclination & 30$^{\circ}$ & 10 \cr
Tully's Group & 32-4+4 & 3 \cr
Nearby Companions within 10 D$_{25}$ & 0 & 11 \cr
Nearby Companions between 10 D$_{25}$ and 20 D$_{25}$ & 1 & 11 \cr
\hline
\end{tabular}
\end{flushleft}
References: 1) Sandage \& Tamman (1981); 2) Garcia-Barreto et al. (2002); 3) Tully (1988); 
4) Veron-Cetty \& Veron (1986,1996); 5) Huchtmeier \& Seiradakis (1985); 6) Soifer et al. (1989); 
7) $\dot M_{OBA}\sim 7.7\times10^{-11} L_{TOT}/L_{\odot} M_{\odot} yr^{-1}$ [Scoville \& Young 
(1983)]; for NGC 3367, $L_{TOT}=L_B+L_{FIR}$$L_{TOT}\sim6.8\times10^{10} L_{\odot}$ ;
8) Grosb\"{o}l (1985); 9) Garcia-Barreto et al. (1993,1996b); 10) Garcia-Barreto \& Rosado (2001);
11) Garcia-Barreto, Carrillo \& Vera-Villamizar (2003)
\end{table}
\clearpage
\newpage

\begin{table}
\small
\caption[ ]{CO(1-0) Parameters}
\begin{flushleft}
\begin{tabular}{lc}
\hline
Phase center $\alpha$ (J2000) & 10$^h$ 46$^m$ 34$^s$.95 \cr
Phase center $\delta$ (J2000) & +13$^{\circ}$ 45$'$ 3$''$ \cr
FWHM low resolution & $4''.26 \times 2''.71$, PA=176$^{\circ}$ \cr
FWHM high resolution & $2''.63 \times 1''.94$ PA=96$^{\circ}$ \cr
Primary beam diameter & 10.4 m \cr
Primary beam (FWHM)  & $53''$ at 114.1 GHz \cr
1 $\sigma$ rms noise in channel maps & 6 mJy / beam \cr
1 $\sigma$ rms noise in high resolution maps & 150 mJy / \kms \cr
1 $\sigma$ rms noise in low resolution maps & 200 mJy / \kms \cr
M(H$_2$) peak ($2''.5 \times 1''.4$) & $3\times 10^8$ M$_{\odot}$ \cr
M(H$_2$) within r=$4''.5$ (r$\leq 950$ pc) & $5.9\times 10^8$ M$_{\odot}$ \cr
M(H$_2$) within r=$27''$ (r$\leq 5.7$ kpc) & $2.6\times 10^9$ M$_{\odot}$ \cr
H$_{2}$ surface density at $R \leq 950$ pc & $\sim 210$ M$_{\odot}$ pc$^{-2}$ \cr
H$_{2}$ surface density at $R \leq 200$ pc & $\sim2.5\times10^3$ M$_{\odot}$ pc$^{-2}$ \cr
Beam average N(H$_2)$  at $R \leq 200$ pc & $\sim1.5\times10^{23}$ cm$^{-2}$ \cr
Beam average N(H$_2)$  at $R \leq 950$ pc & $\sim1.3\times10^{22}$ cm$^{-2}$ \cr
Beam average n(H$_2)$ [N(H$_2) / l; l\sim50$ pc] & $\sim 950$ cm$^{-3}$ \cr
Beam average n(H$_2)$ [N(H$_2) / l; l\sim50$ pc] & $\sim 85$ cm$^{-3}$ \cr
Visual Extinction$^a$, A$_v(1)$, towards nucleus & 150 mag\cr
Star Formation Rate$^b$,  & $\sim 5 M_{\odot} yr^{-1}$ \cr
\hline
\end{tabular}
\end{flushleft}
{\bf (a)} N(H$_2$)/A$_v=0.94\times10^{21}$ cm$^{-2}$ mag$^{-1}$ (\cite{you82}); 
{\bf (b)} $\sim 8\times10^{-11} L_{TOT}$ (\cite{sco83}).
\end{table}
\clearpage
\newpage

\section{FIGURE CAPTIONS}
 
 Figure 1. Spectrum of CO(1-0) emission in NGC 3367 from low resolution (fwhm 4''.2$\times$2''.7 
beam) data smoothed to 15.9 km/s is shown for a 17$''.5$ (1.8 kpc radius) aperture centered on 
the nucleus in NGC 3367. The systemic velocity of the galaxy, V$_{sys}=\sim3032$ \kms is from H${\alpha}$ 
Fabry-Perot.

 Figure 2. CO (1-0) (FWHM 4.2$\times$2.7\arcsec beam) channel maps (twenty channels covering 300 \kms) 
in contours superimposed on the optical continuum I image (Garcia-Barreto 
et al. 1996a,1996b). Contours are at -3,3,4,5,6,7,8,9,10,15,20 $\times$ 6 mJy/beam.

 Figure 3. CO(1-0) integrated line flux over 300 \kms with low angular resolution (beam FWHM $4''.2 \times 2''.7$) 
superimposed on the greyscale H${\alpha}$+[NII] image (Garcia-Barreto et al. 1996a,1996b). CO contours are at -3, 3, 5, 7, 9, 12, 14, 
16, 18, 20, 22, 24, 26, 28, 
30, 35, 40, 45, 50, 60, 70 $\times$ 150 mJy \kms. The weak 
emission structure aligned north-south in the western side of the galaxy seems to originate from beyond 
the stellar bar. The unresolved source seems to originate from the expected location of dust lane in the stellar 
bar, that is, from the north-western side of the stellar bar assuming the bar is rotating counter clockwise (with 
the assumption of trailing spiral arms). 
 
 Figure 4. CO(1-0) integrated line flux over 300 \kms with low angular 
 resolution (beam FWHM $4.2 \times 2.7$\arcsec) 
superimposed on the optical continuum greyscale $I$ image (Garcia-Barreto et al. 1996a,1996b). Contours are at -3, 3, 5, 7, 
9, 12, 14, 16, 18, 20, 22, 24, 26, 28, 30, 
35, 40, 45, 50, 60, 70 $\times$ 150 mJy \kms. Weak emission is 
detected from the NNW and SNE sides of the ends of the stellar bar.
 
 Figure 5. High angular (FWHM $2.6\times1.9$\arcsec) resolution CO(1-0) integrated line flux over 300 \kms (in 
grey scale and contours). Peak of the CO emission is at $\alpha(J2000)=10^h46^m34^s.948, 
\delta(J2000)=+13^{\circ}45'02.7$\arcsec. Secondary peak to the south-west is at $\alpha(J2000)=10^h46^m34^s.66, 
\delta(J2000)=+13^{\circ}45'0.90.$\arcsec Contours are at -3, 3, 5, 7, 9, 12, 14, 16, 18, 
22, 26, 30, 35, 40, 45, 50, 55, 60, 70, 75 $\times$ 150 mJy \kms. The main CO peak position is to be compared with the peak of the 
high resolution synchrotron emission at $\alpha(J2000)=10^h46^m34^s.956, 
\delta(J2000)=+13^{\circ}45'02''.94$ (see Fig. 10).

 Figure 6. CO centroid velocities in contours superposed on 
	the integrated CO line flux in grayscale. The CO systemic velocity is V$_{sys}\sim3035$ km s$^{-1}$. 

 Figure 7. Dispersion velocity map (moment 2) from the central region of NGC 3367 using the low resolution 
map (grey scale and contours). Large velocity dispersion is detected in the central region in PA$\-40^{\circ}$ 
which corresponds to the kinematical minor axis of NGC 3367. Dispersion velocity of 
$\sigma_v\sim40$/sin$i$ km s$^{-1}$ is detected from the NW which would correspond to the crossing of the $x_1$ and 
$x_2$ orbits from the main stellar bar. The largest dispersion, however, is $\sigma_v\sim75$/sin$i$ km s$^{-1}$ in the 
region $\alpha=34^s.66; \; \delta=45'00''.90$ which corresponds to a region with weak or no radio continuum 
synchrotron emission (see Fig. 11 this paper and Fig. 2 in Garcia-Barreto, Franco \& Rudnick 2002).

 Figure 8. Position-velocity diagram from the high resolution map (beam 
 $2''.6\times1''.9$) obtained by adding 
the emission within $\pm2''$ of the major axis (PA=51$^{\circ}$). 
The distance between two ticks along the major axis corresponds to 735 kpc. 
The blue shifted velocity (primary) peak is at 3022 km s$^{-1}$ while the red shifted velocity 
(secondary) peak is at 3068 km s$^{-1}$; their separation is 110 pc. 

 Figure 9. Reproduction of the innermost 3$''$ radio continuum image at 8.4 GHz (in grey scale and contours).
Peak of the radio continuum emission (at an angular resolution of $0''.28\times0''.25$) is at 
$\alpha(J2000)=10^h46^m34^s.956, \delta(J2000)=+13^{\circ}45'02''.94$ (Garcia-Barreto et al. 2002). This 
position is compared to the position of the high resolution CO peak of emission at $\alpha(J2000)=10^h46^m34^s.948, 
\delta(J2000)=+13^{\circ}45'02''.7$ (see Fig. 5)

 Figure 10. Reproduction of the innermost 7$''$ radio continuum image with 1.4 GHz (in contours) and 8.4 GHz (in 
grey scale, Garcia-Barreto et al. 2002)

 Figure 11. Reproduction of the large scale radio continuum image with optical $I$ optical continuum emission (in 
contours) and 1.4 GHz radio continuum emission (in grey scale, \cite{gar02}). Notice the large scale lobes 
straddling the center at a projected radius of 6 kpc.
 

\clearpage
\newpage

\begin{figure}[tbh]
\psfig{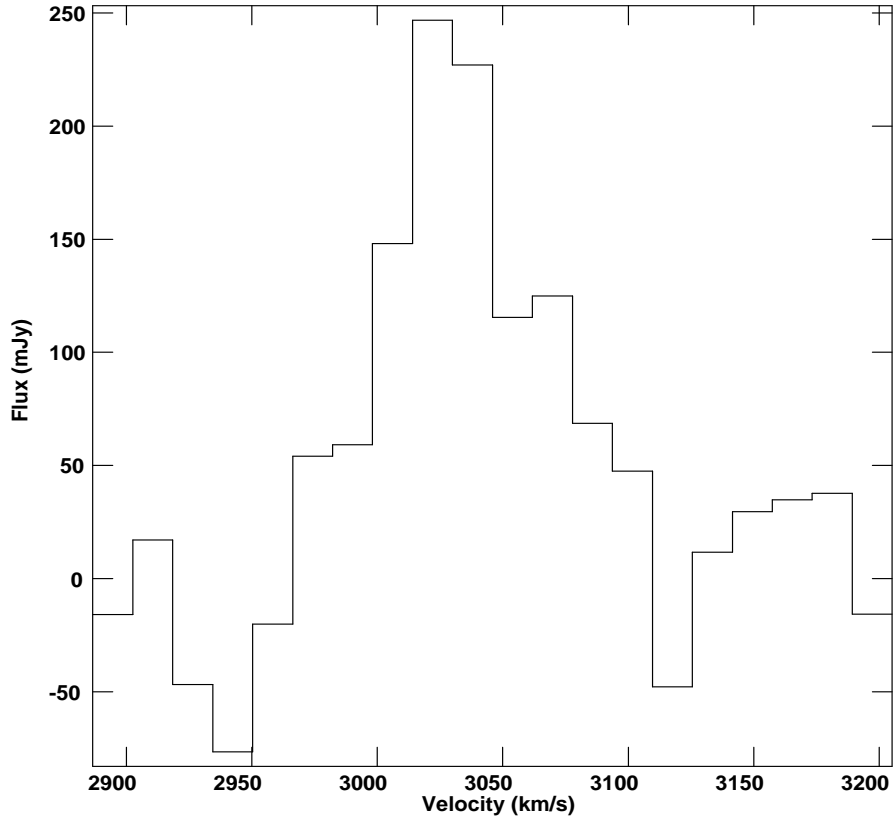}
\caption[]{Spectrum of CO(1-0) emission in NGC 3367 from low resolution (fwhm 4''.2$\times$2''.7 
beam) data smoothed, in velocity, to 15.9 km/s is shown for a 17$''.5$ (1.6 kpc radius) aperture centered on 
the nucleus in NGC 3367. The systemic velocity of the galaxy, V$_{sys}=\sim3032$ \kms is from H${\alpha}$ 
Fabry-Perot (\cite{gar01}).}
\label{fig1}
\label{hi}
\end{figure}
\clearpage
\newpage

\begin{figure}[tbh]
\psfig{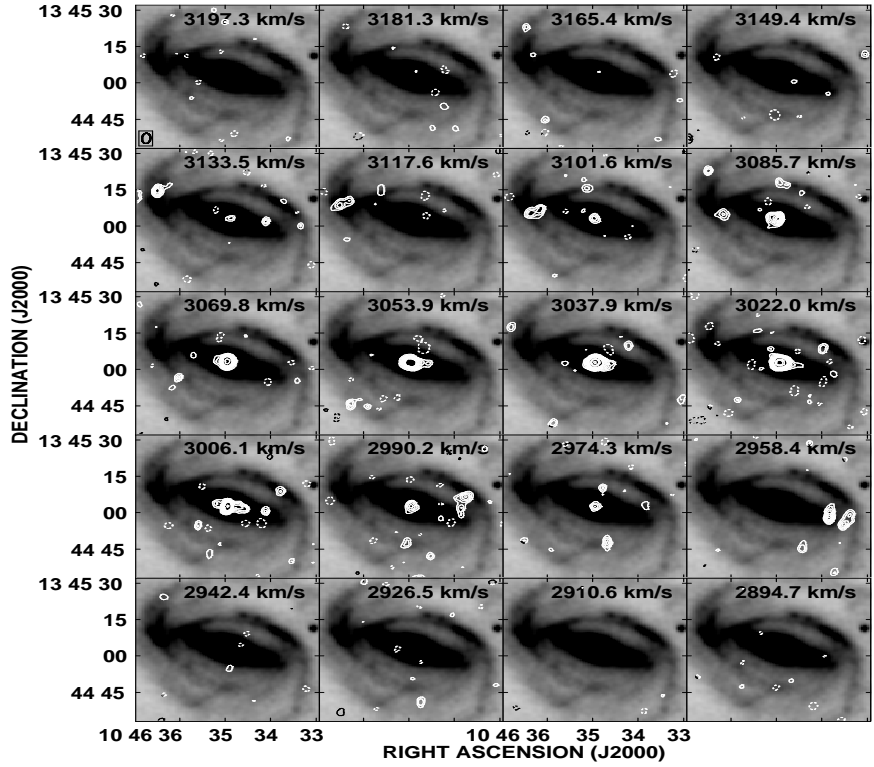}
\caption[]{CO (1-0) (fwhm 4''.2$\times$2''.7 beam) channel velocity maps (twenty channels covering about 300 \kms) 
in contours superimposed on the optical continuum I image (Garcia-Barreto et al. 1996a,1996b). Contours are in units 
of 6 mJy/beam and are -3,3,4,5,6,7,8,9,10,15,20.}
\label{fig2}
\label{hi}
\end{figure}
\clearpage
\newpage

\begin{figure}[tbh]
\psfig{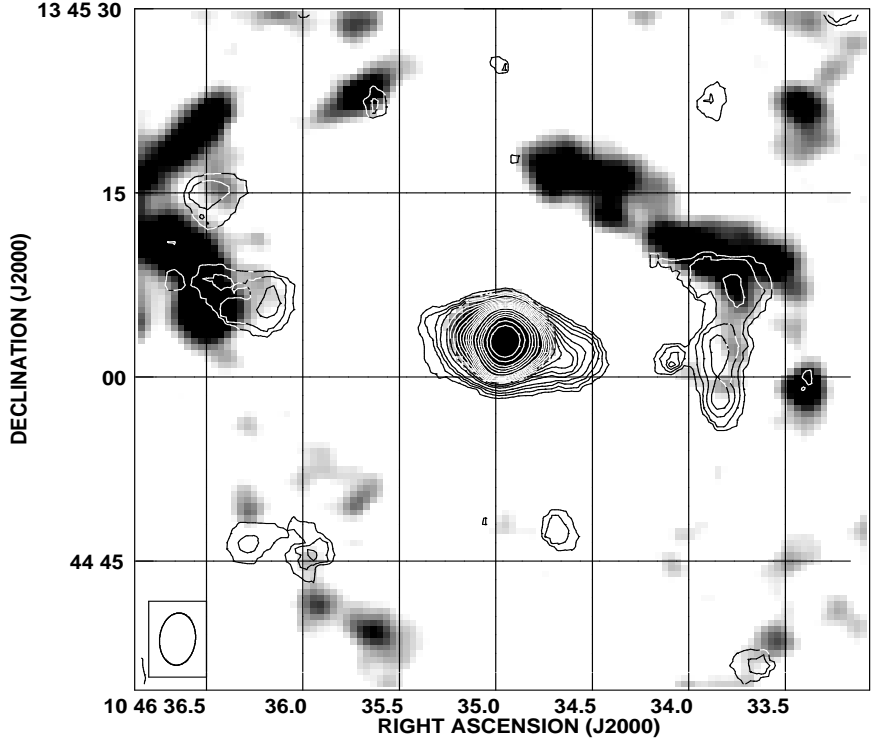}
\caption[]{CO(1-0) integrated line flux over 300 \kms with low angular resolution (beam FWHM $4''.2 \times 2''.7$) 
superimposed on the greyscale H${\alpha}$+[NII] image (Garcia-Barreto et al. 1996a,1996b). CO contours are in units of 
0.15 Jy \kms and are -3, 3, 5, 7, 9, 12, 14, 16, 18, 20, 22, 24, 26, 28, 30, 35, 40, 45, 50, 60, 70. The weak 
emission structure aligned north-south in the western side of the galaxy seems to originate from beyond 
the stellar bar. The unresolved source seems to originate from the expected location of dust lane in the stellar 
bar, that is, from the north-western side of the stellar bar assuming the bar is rotating counter clockwise (with 
the assumption of trailing spiral arms).}
\label{fig3}
\label{hi}
\end{figure}
\clearpage
\newpage

\begin{figure}[tbh]
\psfig{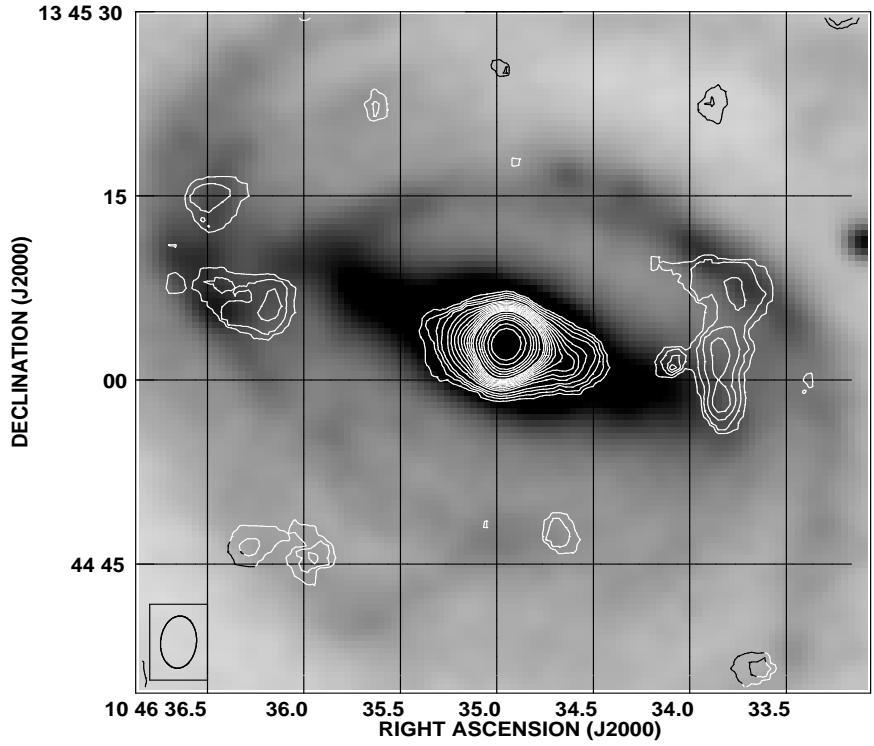}
\caption[]{CO(1-0) integrated line flux over 300 \kms with low angular resolution (beam FWHM $4''.2 \times 2''.7$) 
superimposed on the optical continuum greyscale $I$ image (Garcia-Barreto et al. 1996a,1996b). Contours are in units of 0.15 
Jy \kms and are -3, 3, 5, 7, 9, 12, 14, 16, 18, 20, 22, 24, 26, 28, 30, 35, 40, 45, 50, 60, 70. Weak emission is 
detected from the NNW and SNE sides of the ends of the stellar bar.}
\label{fig4}
\label{hi}
\end{figure}
\clearpage
\newpage

\begin{figure}[tbh]
\psfig{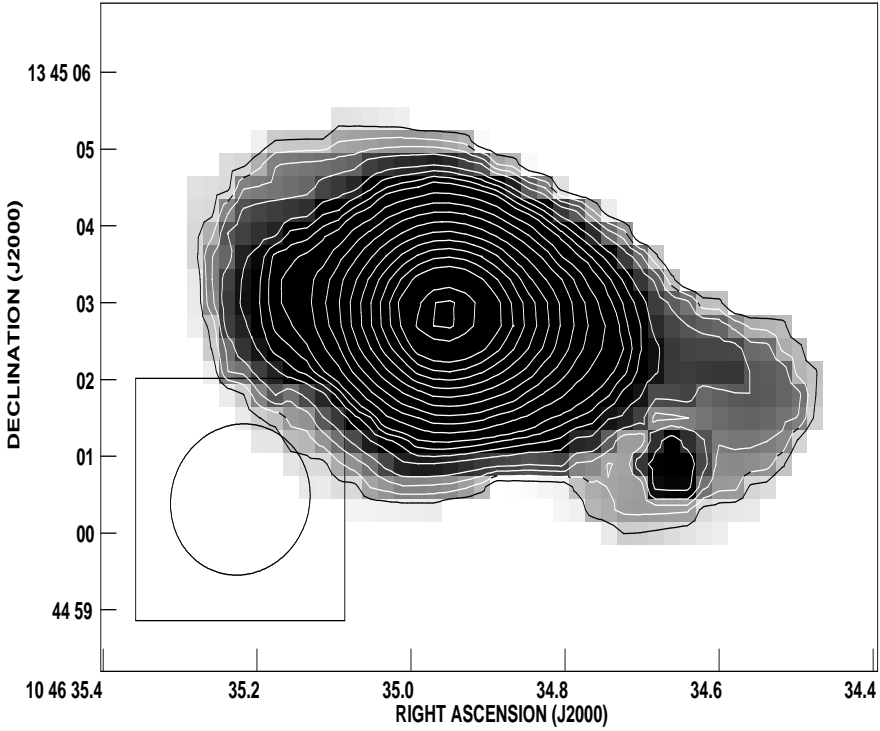}
\caption[]{High angular (FWHM $2''.6\times1''.9$) resolution CO(1-0) integrated line flux over 300 \kms (in 
grey scale and contours). Peak of the CO emission is at $\alpha(J2000)=10^h46^m34^s.948, 
\delta(J2000)=+13^{\circ}45'02''.7$. Secondary peak to the south-west is at $\alpha(J2000)=10^h46^m34^s.66, 
\delta(J2000)=+13^{\circ}45'0''.90.$ Contours are in units of 0.15 Jy \kms and are -3, 3, 5, 7, 9, 12, 14, 16, 18, 
22, 26, 30, 35, 40, 45, 50, 55, 60, 70, 75. The main CO peak position is to be compared with the peak of the 
high resolution synchrotron emission at $\alpha(J2000)=10^h46^m34^s.956, \delta(J2000)=+13^{\circ}45'02''.94$ 
(see Fig. 10).}
\label{fig5}
\label{hi}
\end{figure}
\clearpage
\newpage

\begin{figure}[tbh]
\psfig{figure=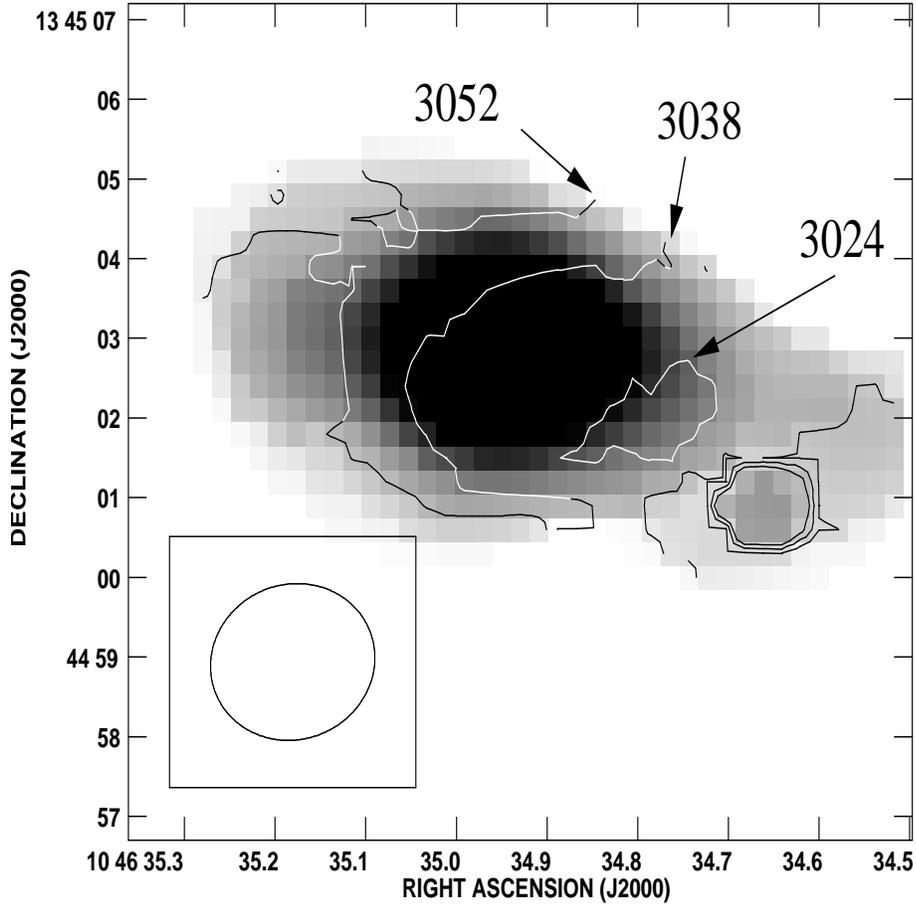,height=12cm,width=12cm,angle=-90}
\caption[]{CO centroid velocities in contours superposed on 
	the integrated CO line flux in grayscale. The CO systemic velocity is V$_{sys}\sim3035$ km s$^{-1}$. }
\label{fig6}
\label{hi}
\end{figure}
\clearpage
\newpage

\begin{figure}[tbh]
\psfig{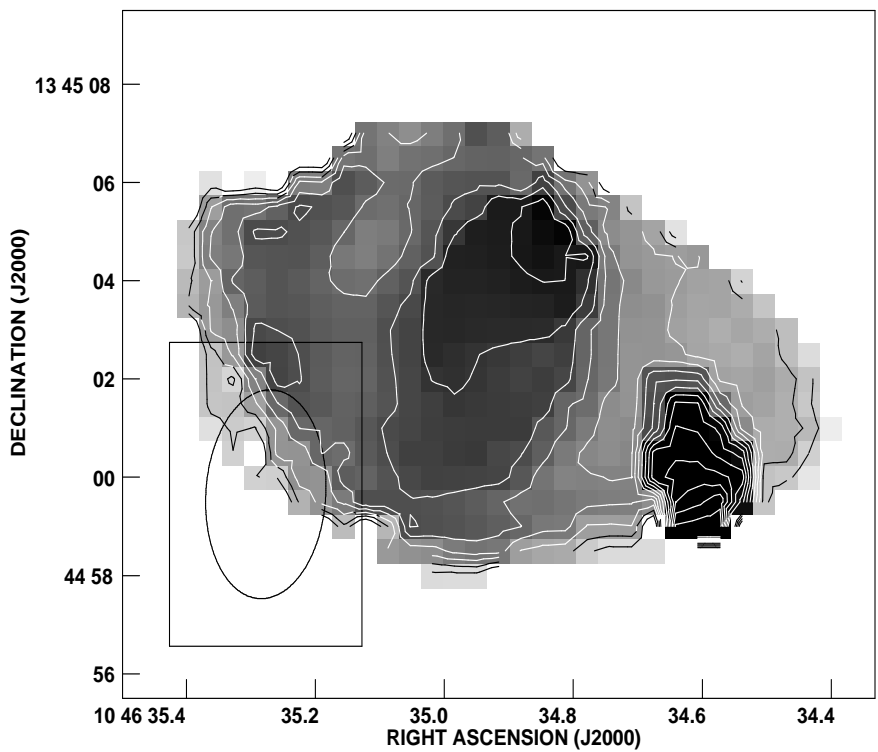}
\caption[]{Dispersion velocity map (moment 2) from the central region of NGC 3367 using the low resolution 
map (grey scale and contours). Large velocity dispersion is detected in the central region in PA$\-40^{\circ}$ 
which corresponds to the kinematical minor axis of NGC 3367. Dispersion velocity of 
$\sigma_v\sim40$/sin$i$ km s$^{-1}$ is detected from the NW which would correspond to the crossing of the $x_1$ and 
$x_2$ orbits from the main stellar bar. The largest dispersion, however, is $\sigma_v\sim75$/sin$i$ km s$^{-1}$ in the 
region $\alpha=34^s.66; \; \delta=45'00''.90$ which corresponds to a region with weak or no radio continuum 
synchrotron emission (see Fig. 11 this paper and Fig. 2 in Garcia-Barreto, Franco \& Rudnick 2002).}
\label{fig7}
\label{hi}
\end{figure}
\clearpage
\newpage

\begin{figure}[tbh]
\psfig{figure=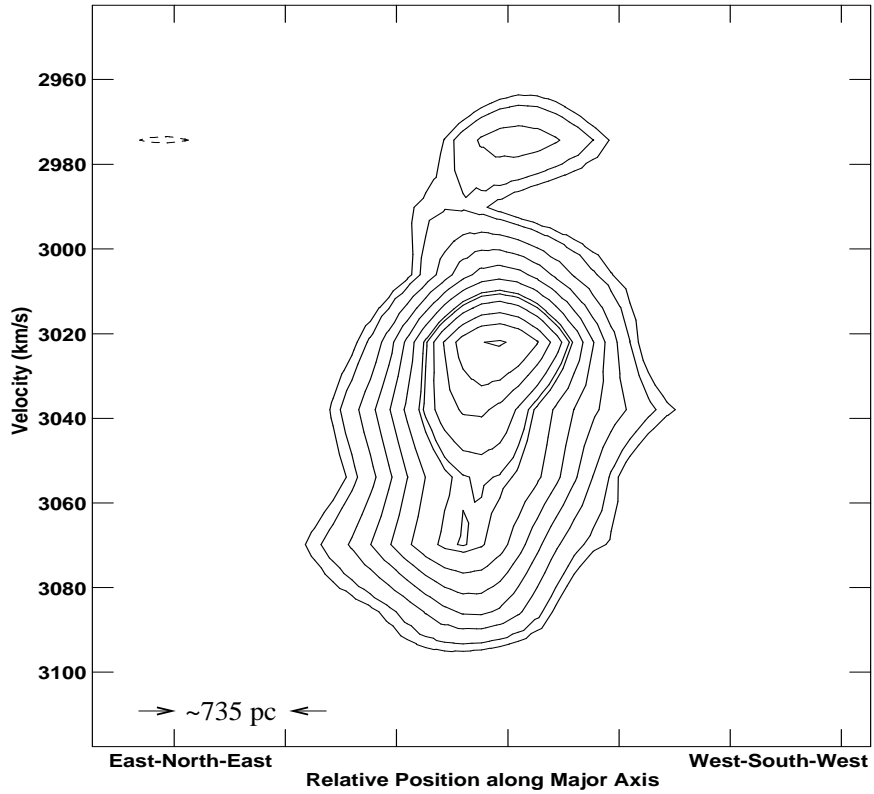,height=12cm,width=12cm}
\caption[]{Position-velocity diagram from the high resolution map (beam 
 $2''.6\times1''.9$) obtained by adding 
the emission within $\pm2''$ of the major axis (PA=51$^{\circ}$). 
The distance between two ticks along the major axis corresponds to $\sim$735 pc. 
The blue shifted velocity (primary) peak is at 3022 km s$^{-1}$ while the red shifted velocity 
(secondary) peak is at 3068 km s$^{-1}$; their separation is 110 pc. }
\label{fig8}
\label{hi}
\end{figure}
\clearpage
\newpage

\begin{figure}[tbh]
\psfig{figure=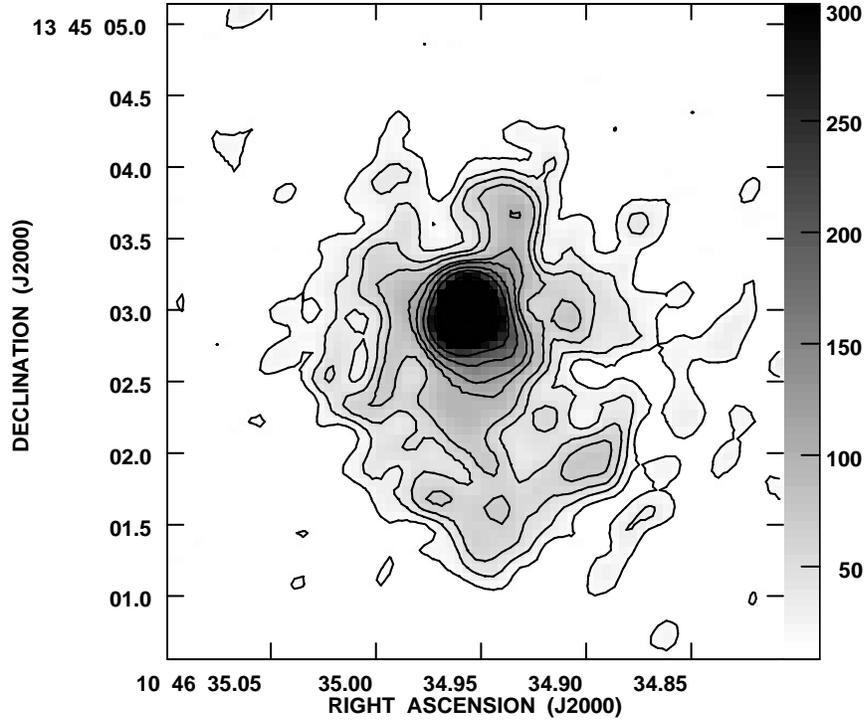,height=12cm,width=12cm}
\caption[]{Reproduction of the innermost 3$''$ radio continuum image at 8.4 GHz (in grey scale and contours).
Peak of the radio continuum emission (at an angular resolution of $0''.28\times0''.25$) is at 
$\alpha(J2000)=10^h46^m34^s.956, \delta(J2000)=+13^{\circ}45'02''.94$ (Garcia-Barreto et al. 2002). This 
position is compared to the position of the high resolution CO peak of emission at $\alpha(J2000)=10^h46^m34^s.948, 
\delta(J2000)=+13^{\circ}45'02''.7$ (see Fig. 5).}
\label{fig9}
\label{hi}
\end{figure}
\clearpage
\newpage

\begin{figure}[tbh]
\psfig{figure=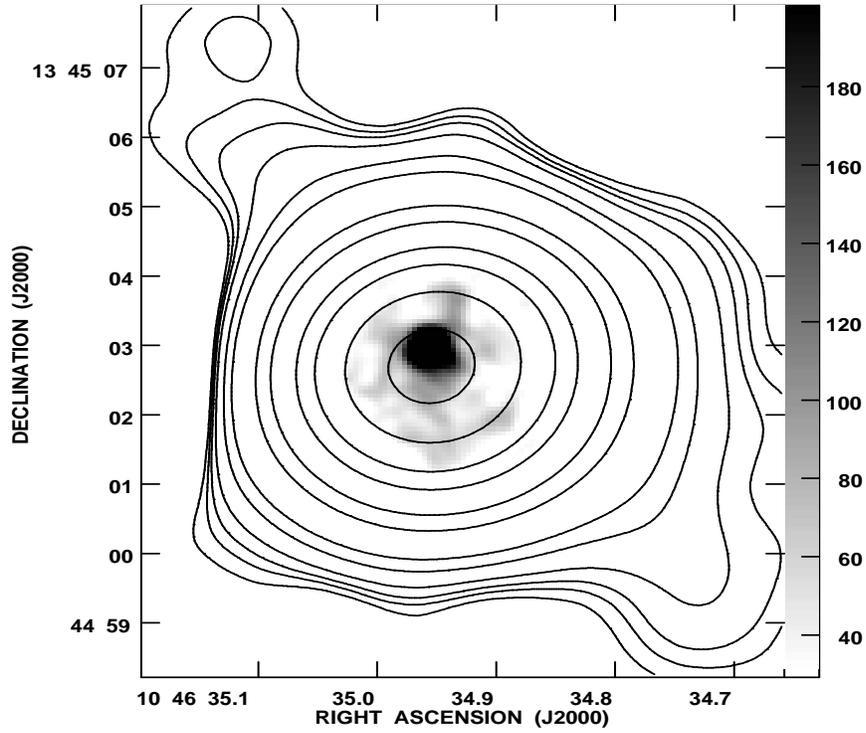,height=12cm,width=12cm}
\caption[]{Reproduction of the innermost 7$''$ radio continuum image with 1.4 GHz in contours and 8.4 GHz in grey scale 
(\cite{gar02})}
\label{fig10}
\label{hi}
\end{figure}
\clearpage
\newpage

\begin{figure}[tbh]
\psfig{figure=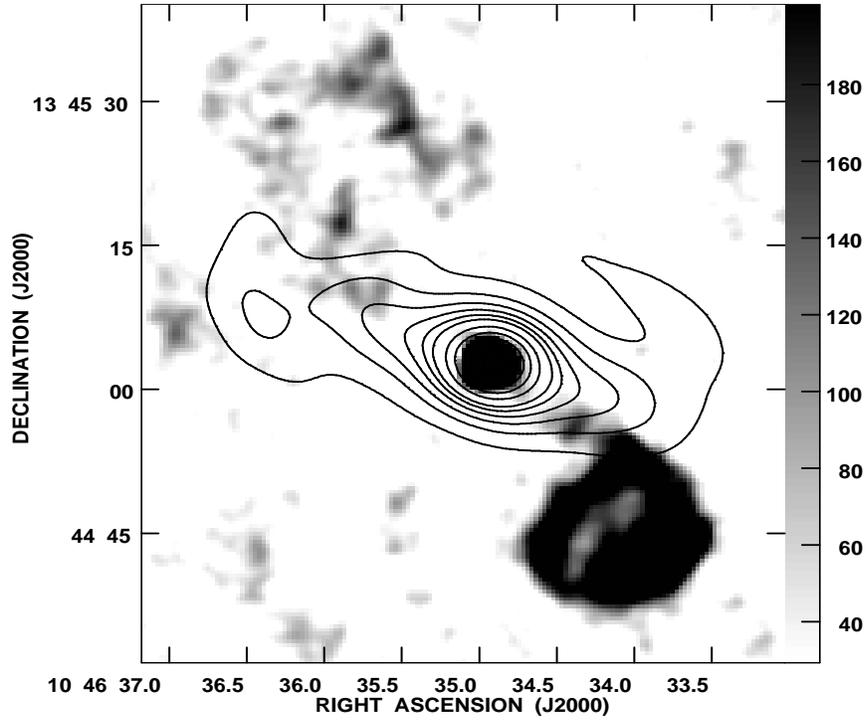,height=12cm,width=12cm}
\caption[]{Reproduction of the large scale radio continuum 1.4 GHz emission image (in grey scale) superposed to 
optical $I$ continuum emission (in contours; \cite{gar02}). Notice the large scale lobes 
straddling the center at a projected radius of 6 kpc.}
\label{fig11}
\label{hi}
\end{figure}

\end{document}